\documentclass[proof]{WileyASNA-v1}

\articletype{Original Article}%

\received{26 April 2016}
\revised{6 June 2016}
\accepted{6 June 2016}

\begin{document}

\title{Extremely inverted peaked spectrum radio sources} 

\author[1]{Mukul Mhaskey*}

\author[2]{Surajit Paul}

\author[3]{Gopal Krishna}

\authormark{MHASKEY \textsc{et al}}

\address[1]{\orgname{Th\"uringer Landessternwarte}, \orgaddress{\state{Sternwarte 5, D-07778 Tautenburg}, \country{Germany}}}

\address[2]{\orgdiv{Department of Physics}, \orgname{Savitribai Phule Pune University}, \orgaddress{\state{Ganeshkhind, Pune 411007}, \country{India}}}

\address[3]{\orgname{ UM-DAE Centre for Excellence in Basic Sciences (CEBS)}, \orgaddress{\state{University of Mumbai Campus, Vidya Nagari, Mumbai 400098}, \country{India}}}

\corres{*Mhaskey Mukul, Th\"uringer Landessternwarte, Tautenburg, Germany. \email{mukul@tls-tautenburg.de}}


\abstract{We report our ongoing search for extremely inverted spectrum compact radio galaxies, for which the defining feature in the radio spectrum is not the spectral peak, but instead the slope of the spectrum (alpha) in the high-opacity (i.e., lower frequency) part of the radio spectrum. Specifically, our focus is on the spectral regime with spectral index, $\alpha_{thick} > $+2.5. The motivation for our study is, firstly, extragalactic sources with such extreme spectral index are extremely rare, because of the unavailability of right combination of sensitivity and resolution over a range of low frequencies. The second reason is more physically motivated, since alpha = +2.5 is the maximum slope theoretically possible for a standard radio source emitting synchrotron radiation. Therefore such sources could be the test-bed for some already proposed alternative scenarios for synchrotron self-absorption (SSA), like the free-free absorption (FFA) highlighting the importance of jet-ISM interaction in the radio galaxy evolution.} 



\jnlcitation{\cname{%
\author{Mukul Mhaskey}, 
\author{Surajit Paul}, and 
\author{Gopal-Krishna}} (\cyear{2021}), 
\ctitle{Extremely inverted peaked spectrum radio sources}, \cjournal{AN}, 
\cvol{2021;00:x--x}.}


\maketitle


\section{Introduction}\label{sec1}
Gigahertz peaked spectrum (GPS) and compact steep spectrum (CSS) sources are compact radio-loud active galaxies with linear sizes of a few kilo-parsecs. They have a characteristic inverted radio-continuum spectrum which is a result of absorption of the synchrotron radiation at giga-hertz ($\nu_{peak} < 1.0$ GHz, GPS) and sub-GHz ($\nu_{peak} < 0.5$ GHz, CSS) frequencies, caused either by synchrotron self absorption (SSA) or free-free absorption (FFA), or a combination of both \citep[e.g.,][]{Odea1998}. GPS/CSS sources offer a unique opportunity to study the AGN-host galaxy feedback process because the radio emission is embedded within their host galaxies. Furthermore, there is evidence to suggest that these sources are associated with new or recent AGN activity, and thus are ideal candidates to study jet evolution and galaxy formation \citep{Odea1998, Odea2021}. This has however been contested due to the large numbers of GPS/CSS sources found in radio surveys and favours the alternative scenario in which the radio emission is associated with an old AGN activity and the emission cannot escape the host galaxy due to unusually dense environment. The absorption mechanisms that give rise to the inverted spectrum can resolve this ambiguity about the nature of the GPS/CSS sources since several physical factors in these sources govern the absorption mechanism. But these absorption mechanisms are still poorly understood because of lack of sufficient flux measurements to sample the spectrum 
\citep{Odea1998, Callingham2017}.

In an attempt to understand the absorption mechanism in such sources and study the environmental interactions between AGN jets and the host galaxy ISM, we started a search program with the GMRT to look for inverted spectrum radio sources that have a spectral index, $\alpha^{\rm 325\, MHz}_{\rm 150\, MHz}$ $>+$2.5 ($S\propto\nu^{-\alpha}$). 
This is larger than the theoretically allowed spectral index limit for SSA in a perfectly homogeneous source which emits synchrotron radiation. So far, we have found 12 sources with $\alpha$ near $+$2.5 We have named such sources as `Extremely Inverted Spectrum Extragalactic Radio Sources' \citep[EISERS;][]{Mhaskey2019a,Mhaskey2019b, Mhaskey2020}.


\section{Sample}\label{sec3}
The two main radio surveys, viz., the TGSS-ADR1 (150 MHz) and WENSS (325 MHz), overlap a region of about 1.03$\pi$ steradian ($\sim$1/4$^{th}$ of the entire sky). The selected overlapping region contains 229420 radio sources in the WENSS catalogue covering the declination north of +28$^\circ$. Out of this extensive source list, we first extracted a subset of 35064 sources which belong to the morphological type \lq S\rq~(i.e., single, as per the WENSS catalogue) and are stronger than 150 mJy at 325 MHz. For each of these shortlisted WENSS sources, we then looked for a TGSS-ADR1 counterpart, within a search radius of 20 arcsec. We found counterparts for 33707 (out of 35064) of the WENSS sources. For 1357 WENSS sources no counterparts in TGSS-ADR1 were found. 
The spectral index, $\alpha$, between 150 and 325 MHz, for all the sources was calculated. For the WENSS sources without a counterpart (1357) in the TGSS-ADR1, a 5-$\sigma$ limit to the flux at 150 MHz was used to estimate the upper limits for the spectral index. Here $\sigma$ is the image rms for individual sources in TGSS-ADR1.    

The final list of EISERS candidates satisfy the following criteria: (i) structural type \lq S\rq~and flux density $>$ 150 mJy at 325 MHz, (ii) omission of sources found to lie within 10$^{\circ}$ of the galactic plane, or listed as H-II regions in the NASA Extragalactic Database (NED)\footnote{https://ned.ipac.caltech.edu/} (iii) a clean detection in the WENSS (325 MHz) and NVSS (1.4 GHz) (based on visual inspection of the respective radio images) and (iv) $\alpha^{325\;MHz}_{150\;MHz} >$ $+$2.75, in the case of sources detected in TGSS-ADR1 or $\alpha^{325\;MHz}_{150\;MHz} >$ $+$2.5, in the case of sources not-detected in TGSS-ADR1. A conservative threshold value of +2.75 was adopted for sources detected at 150 MHz in order to make an allowance for flux variability. This could be significant in the case of compact sources \citep{Bell2018,Chhetri2018}. The final list contains 29 EISERS candidates which were later observed quasi-simultaneous in band 2, 3 and 4 with the uGMRT.

\section{Observations and data analysis}\label{sec4}
The need for quasi-simultaneous radio observations at frequencies below the spectral turnover is underscored by the fact that the two radio surveys (TGSS-ADR1 and WENSS) used for computing the spectral slopes of these compact radio sources had been made nearly a decade apart. The long time interval could then have introduced significant uncertainty due to flux variability expected from refractive interstellar scintillation at such low frequencies, e.g. \citep{Bell2018}. It may also be noted that the WENSS \citep{Rengelink1997} is known to be off the flux-density scales defined by Roger, Costain \& Bridle \citep[RCB,][]{Roger1973} and by \citet{Baars1977} by over 10\% \citep[see,][]{Hardcastle2016}. For some areas in the sky the TGSS-ADR1 reports systematically low fluxes, sometimes even by 40-50\% , also $\sim$5-10\% of the survey area has systematic flux deviations of $>$10\% \citep{Intema2017}. Therefore, the previous spectral index estimates could be substantially inaccurate due to the combined effect of measurement uncertainty and the calibration uncertainties of the WENSS and TGSS-ADR1 flux densities. This may account for the substantial differences found in some of the cases, between the WENSS flux density at 325 MHz and the present uGMRT measurements at the same frequency. For our sample of EISERS candidates, the present GMRT observations have yielded the data with the highest sensitivity and resolution currently available at 150 and 325 MHz. Moreover, the availability of two data points at well-spaced frequencies in their highly opaque spectral region raises the confidence in quantifying the steepness of the spectral turnover. 

The 29 EISERS candidates were observed with the recently upgraded GMRT \citep[`uGMRT',][]{Gupta2017}. The observations were performed in a snapshot mode, quasi-simultaneously at bands 2, 3 and 4, using the wide band receivers covering a frequency range from 150 to 900 MHz. The wide band of the uGMRT was divided into several smaller sub-bands and then imaged separately. The measured visibilities at 150 MHz and 325 MHz were processed using the Source Peeling and Atmospheric Modelling \citep[\textsc{spam}][]{Intema2014} package. \textsc{spam} is a semi-automated pipeline based on \textsc{aips}, \textsc{parseltongue} and \textsc{python}. It performs a series of iterative flagging and calibration sequences and the imaging involves direction dependent calibration. This package has been used for processing of the entire TGSS data at 150 MHz \citep{Intema2017}. Details of \textsc{spam} and its various routines are provided in \citet{Intema2017}. Thus several flux measurements between 150 and 900 MHz were obtained (see Figure 1).

Out of the 29 candidates observed 12 have a very steep spectral index with $\alpha^{325\;MHz}_{150\;MHz} >$ $+$2.0. In Table 1 the GMRT flux densities found here at 150 and 325 MHz are listed together with the measured spectral index. This is the final list of confirmed EISERS. 

\begin{table}[h!]
\centering
\caption{Source name, flux densities and spectral indices (150-325 MHz, uGMRT) of the extremely inverted spectrum sources.}
\begin{tabular}{cccc}
\hline

\multicolumn{1}{c}{Source}& \multicolumn{2}{c}{uGMRT flux density (mJy)} & \multicolumn{1}{c}{Spectral index} \\
  &  150 MHz & 325 MHz & $\alpha_{\rm 150\,MHz}^{\rm 325\,MHz}$\\
\hline
J0304+7727 & 32.2${\pm}3.7$ &167.2${\pm}$16.7 &2.1${\pm}$0.2\\
J0614+6108 &29.1${\pm}$5.1 &146.9${\pm}$14.7 &2.1${\pm}$0.3\\
J0649+5947 &21.4${\pm}$4.3 &128.1${\pm}$12.8 &2.3${\pm}$0.3\\
J0740+7129 &12.2${\pm}$2.8 &095.6${\pm}$09.6 &2.7${\pm}$0.3\\
J0847+5723 &29.1${\pm}$4.8 &180.9${\pm}$18.1 &2.4${\pm}$0.3\\
J0853+6722 &17.5${\pm}$3.1 &087.7${\pm}$08.8 &2.1${\pm}$0.3\\
J0858+7501 &25.4${\pm}$6.1 &162.2${\pm}$16.2 &2.4${\pm}$0.3\\
J1326+5712 &11.5${\pm}$1.4 &108.9${\pm}$10.9 &2.9${\pm}$0.2\\
J1536+8154 &35.1${\pm}$5.9 &198.9${\pm}$19.9 &2.2${\pm}$0.3\\ 
J1549+5038 & 44.3${\pm}$6.2&287.8${\pm}$28.8 &2.4${\pm}$0.2\\
J1658+4732 &15.4${\pm}$2.8 &163.4${\pm}$16.4 &3.0${\pm}$0.3\\
J2317+4738 &29.7${\pm}$3.7 &145.0${\pm}$14.5 &2.1${\pm}$0.2\\
\hline
\end{tabular}
\label{table:spec-prop}
\end{table}



\section{Discussion and Conclusion}\label{sec5}
The  main  objective of this project is to understand the jet-ISM environment  in  GPS/CSS  sources by  probing  the synchrotron radiation absorption medium and to distinguish between the SSA and FFA mechanisms. For this we concentrate our efforts on sources with extremely inverted radio spectra, since they are the sites of maximum synchrotron radiation opacity.
We conducted a first systematic search for the rare sub-class of inverted spectrum extragalactic radio sources which show an inverted (integrated) radio spectrum with a spectral index $\alpha >$ +2.0. 

These sources seem to be particularly interesting from the viewpoint of violation of the theoretical SSA limit in compact extragalactic radio sources and may thus require alternative explanations. 
One possibility is that the energy spectrum of the relativistic particles in EISERS deviates from the canonical shape (a power-law), to a Maxwellian or mono-energetic particle energy distribution \citep{Rees1967}.

\begin{figure}
\begin{center}
\includegraphics[trim={1.75cm 0cm 2.0cm 0.5cm},clip,width=0.45\textwidth]{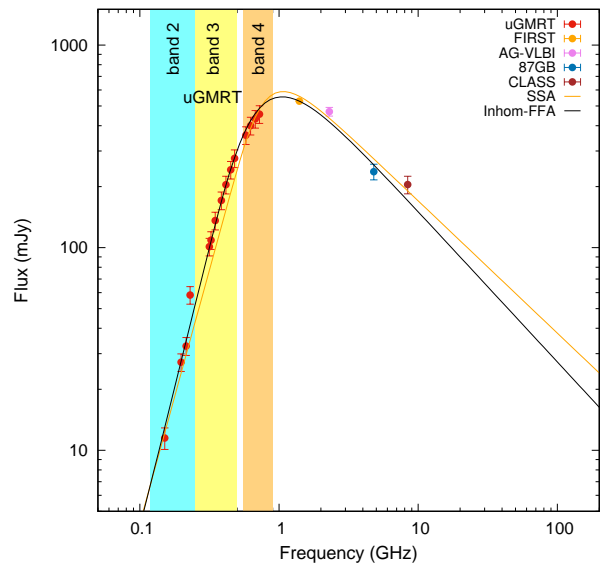}
\end{center}
\caption{\scriptsize{Radio spectrum of J1326+5712. The filled red circles represent the measurements from the uGMRT observations at the multiple narrow frequency sub-bands in band 2, band 3 and band 4. The shaded regions represents the range of frequencies for different bands observable by the uGMRT. The SSA and FFA absorption models are represented by the yellow and black curves respectively.}} \label{fig:spec_all} 
\end{figure}

An alternative explanation for EISERS invokes free-free absorption (FFA) \citep{Kellermann1966, Bicknell1997, Kuncic1998}. As mentioned above, EISERS violate the standard synchrotron picture. We have modelled the radio spectra of the confirmed EISERS in terms of an inhomeogeneous FFA screen external to the radio source, following the prescription of \citet{Bicknell2018}. Earlier, they had suggested that such a thermal screen can arise, as the bow shock produced by the advancing jet photo-ionises the thermal gas clouds filling the ambient (interstellar) space \citep{Bicknell1997}. A satisfactory fit to the radio spectrum was found for a majority of the confirmed EISERS. This is demonstrated for the source J1326+5712 ($\alpha^{325\;MHz}_{150\;MHz} =$ $+$2.91$\pm$0.2) in Figure 1, where the curve representing the the FFA model is a better fit than the curve for the SSA model. Further results are published in \citet{Mhaskey2019b}. The derived estimates of the mean electron density within the absorbing gas were found to match well with the densities predicted in the theoretical models for jet propagation in the ionised medium, by \citet{Begelman1996} and \citet{Bicknell1997}. Any evidence for high electron densities \citep[n$_{e}  >$100\,cm$^{-3}$,][]{Bicknell2018} in EISERS would be consistent with the FFA explanation. 


Also, the simulations of relativistic jets interacting with a warm, inhomogeneous medium \citep{Mukherjee2016,Bicknell2018} show that free-free absorption can account for the $\sim$GHz peak frequencies and low frequency power-laws inferred from the radio observations. 
At early stages, the low frequency spectrum is steep but progressively flattens as a result of a broader distribution of optical depths. 
Since EISERS have the steepest of slopes below the turnover in the optically thick regime, it can be assumed that these sources are very young AGNs and would eventually evolve into normal radio galaxies.

Our initial results indicate that the inhomogeneous-FFA model by \citep{Bicknell1997} is  typically preferred  over  single,  homogeneous SSA and FFA absorbing media.  To pin down the absorption mechanism with high statistical  confidence,  more  data  points  both  in  the  optically thick and thin part of the spectrum below and above the turnover frequency are crucial.  In order to  assess  the  impact  an  inhomogeneous  absorbing medium has on the absorbed spectrum, we are numerically building a new model which incorporates both SSA and FFA mechanisms and will compare them with our observations. 

\section*{Acknowledgments}

We thank the staff of the GMRT who have made these observations possible. GMRT is run by the National Centre for Radio Astrophysics of the Tata Institute of Fundamental Research. This research has used NASA's Astrophysics Data System and NASA/IPAC Extragalactic Database (NED), Jet Propulsion Laboratory, California Institute of Technology under contract with National Aeronautics and Space Administration and VizieR catalogue access tool, CDS, Strasbourg, France. SP would like to thank DST INSPIRE Faculty Scheme (IF12/PH-44) for funding his research group. G-K acknowledges the Senior Scientist Fellowship of the Indian National Science Academy.

\bibliography{Wiley-ASNA}%

\section*{Author Biography}
\begin{biography}{}{\textbf{Mukul Mhaskey} obtained his PhD from Savitribai Phule Pune University, Pune, India in Dec, 2020 and is currently a postdoctoral researcher at the Th\"uringer Landessternwarte, Tautenburg, Germany. His main researchinterests are multi-wavelength studies of radio-emitting active galactic nuclei to understand their evolution and the feedback with host galaxies.}
\end{biography}
\end{document}